\documentclass[sigconf]{acmart}
\usepackage{natbib}
\usepackage{array}
\usepackage[utf8]{inputenc}
\usepackage{tabularx}
\usepackage{comment}
\usepackage{rotating}
\newcolumntype{Y}{>{\centering\arraybackslash}X} 

\usepackage{tabularx}
\usepackage{ragged2e} 
\AtBeginDocument{%
  }

\copyrightyear{2026}
\acmYear{2026}
\setcopyright{cc}
\setcctype{by}
\acmConference[CHI EA '26]{Extended Abstracts of the 2026 CHI Conference on Human Factors in Computing Systems}{April 13--17, 2026}{Barcelona, Spain}
\acmBooktitle{Extended Abstracts of the 2026 CHI Conference on Human Factors in Computing Systems (CHI EA '26), April 13--17, 2026, Barcelona, Spain}
\acmDOI{10.1145/3772363.3798378}
\acmISBN{979-8-4007-2281-3/2026/04}

\begin{document}

\title[Conversational Successes and Breakdowns in Everyday Non-Display Smart Glasses Use]{Conversational Successes and Breakdowns in Everyday Smart Glasses Use}

\author{Xiuqi Tommy Zhu}
\email{zhu.xiu@northeatern.edu}
\affiliation{%
  \institution{College of Arts, Media and Design, Northeastern University}
  \city{Boston}
  \state{MA}
  \country{USA}
}

\author{Xiaoan Liu}
\email{xiaoan.liu@colorado.edu}
\affiliation{%
  \institution{University of Colorado Boulder}
  \city{Boulder}
  \state{CO}
  \country{USA}
}

\author{Casper Harteveld}
\email{c.harteveld@northeastern.edu}
\affiliation{%
  \institution{College of Arts, Media and Design, Northeastern Universityn}
  \city{Boston}
  \state{MA}
  \country{USA}
}

\author{Smit Desai}
\orcid{0000-0001-6983-8838}
\email{sm.desai@northeastern.edu}
\affiliation{%
  \institution{Northeastern University}
  \city{Boston}
  \state{MA}
  \country{USA}
}

\author{Eileen McGivney}
\email{e.mcgivney@northeastern.edu}
\affiliation{%
  \institution{College of Arts, Media and Design, Northeastern University}
  \city{Boston}
  \state{MA}
  \country{USA}
  \footnote{Corresponding author}
}

\renewcommand{\shortauthors}{Zhu et al.}

\begin{abstract}
Non-Display Smart Glasses hold the potential to support everyday activities by combining continuous environmental sensing with voice-only interaction powered by large language models (LLMs). Understanding how conversational successes and breakdowns arise in everyday contexts can better inform the design of future voice-only interfaces. To investigate this, we conducted a month-long collaborative autoethnography (\textit{n}=2) to identify patterns of successes and breakdowns when using such devices. We then compare these patterns with prior findings on voice-only interactions to highlight the unique affordances and opportunities offered by non-display smart glasses.
\end{abstract}




\maketitle
\section{Introduction}
Integrating Large Language Models (LLMs)~\cite{Yin_2024} into \textit{non-display smart glasses}\footnote{We use `non-display smart glasses’ to refer to voice-only interaction glasses with multimodal input ability, without augmented reality (AR) displays.} is reshaping everyday activities by enabling hands-free, context-aware, multimodal interaction \cite{waisberg_meta_2024}. These devices move computing beyond the screen, allowing voice assistants (VAs)\footnote{Following prior literature, we use the term VAs interchangeably with conversational agents (CA), virtual personal assistants (VPA), personal digital assistants (PDA), and intelligent personal assistants (IPA) \cite{cowan_what_2017, de_barcelos_silva_intelligent_2020, hoy_alexa_2018}.} to perceive and respond to the environment through speech and visual input \cite{trichopoulos_smart_2024, kim_applications_2021}. For example, the Meta Ray-Ban AI Glasses\footnote{\url{https://www.meta.com/ai-glasses/}} allow users to say “\textit{Hey Meta, look and tell me about…}” and receive responses grounded in first-person visual context—reflecting a broader shift toward \textit{Heads-up Computing} \cite{zhao_heads-up_2023}.

Prior work has demonstrated the importance of understanding conversational successes and breakdowns in everyday voice-only interfaces such as smart speakers \cite{porcheron_voice_2018, bentley_understanding_2018, li_multi-modal_2020, Desai_Twidale_2023}. For example, users often use such VAs successfully for routine tasks, including playing music, setting alarms, and controlling IoT devices \cite{cowan_what_2017, ammari_music_2019, bentley_understanding_2018}, yet recurrent conversational breakdowns, such as misrecognized intent, oversimplified commands, and poor error recovery, remain central sources of frustration \cite{porcheron_voice_2018, mahmood2025user}. These challenges persist even with LLMs-powered VAs, which may still produce vague answers, misinterpret open-ended queries, or terminate tasks prematurely \cite{mahmood2025user, yang_talk2care_2024, chan2025mango}. Such failures often stem from mismatches between user expectations and system capabilities \cite{luger_like_2016}, as well as unmet needs for competence, autonomy, and shared understanding \cite{wang_towards_2021, Desai_Twidale_2023, mahmood2025user}, and may lead users to abandon using VAs. By contrast, non-display smart glasses fundamentally differ from fixed voice-only interfaces in that their interactions are not only triggered by voice but also grounded in first-person visual perception, which introduces new opportunities, and potential failures, arising from shared visual awareness with LLMs \cite{bentley_understanding_2018, cowan_what_2017, porcheron_voice_2018, terzopoulos2020voice}. Given the novelty of this technology, it remains unclear what kinds of conversational successes and breakdowns may arise during everyday smart glasses use. Establishing preliminary use patterns can inform the design of future smart glasses beyond the current non-display limitations\cite{zhao_heads-up_2023}. We aimed to bridge these gaps by answering this core research question:

\textbf{\textit{RQ: How do conversational successes and breakdowns with smart glasses manifest during everyday activities, and what difficulties or impacts do they present for users?}}

Therefore, we conducted a collaborative autoethnography (CAE) \cite{jones2016introduction, lapadat_ethics_2017}, as it allowed us to capture situated experiences of breakdowns and successes in naturalistic settings over one month (\textit{n}=2). We discovered recurring success patterns in supporting instant referential problem-solving, understanding unfamiliar knowledge, and decision-making in everyday uses (Session 3.1). Further, we identified recurring breakdown patterns, including referential incoherence, conflicts with human perception, social embarrassment, and voice-only interaction paradigm limitations (Session 3.2).
We discuss how these patterns are unique in that non-display smart glasses effectively support queries involving ambiguous references (e.g., ``this one'), but often struggle to maintain referential coherence due to misalignments between user intent and system understanding.

\section{Collaborative Autoethnography}
As an emerging technology, non-display smart glasses are still rarely integrated into everyday life, and little is known about how they are experienced in naturalistic contexts. Thus, an autoethnographic approach is particularly well-suited to capture the exploratory, first-hand, and evolving nature of these interactions. To mitigate individual bias, we adapt Collaborative Autoethnography (CAE), which emphasizes collective reflection, enabling personally engaging and non-exploitative research \cite{jones2016introduction, lapadat_ethics_2017}. Similar to duoethnography \cite{neupane_wearable_2025} and trioethnography \cite{desai_using_2023}, it situates researchers in dialogical relationships that support the (re)construction and reflection of personal narratives \cite{kaltenhauser_playing_2024}. This dialogic process allows HCI researchers (us) to collaboratively analyze our \textit{“pooled autoethnographic data as a collective”} \cite{wakkary_backyard_2025, ciolfi_felice_doing_2025}. 

\subsection{Positionality Statement}
To ground this study, we shared our positionalities with the research team \cite{singh_exploring_2025}. The first author [T] is an interdisciplinary doctoral student whose work centers on AR/VR, human–AI collaboration, and educational computing within the broader field of HCI. The second author [X] is a doctoral student working in spatial computing and augmented reality, prioritizing pragmatic usability and cultivating an optimistic yet critical view of wearable AI, which shapes his interpretation of the study data. As both international asian students, Mandarin is their first language, and English is second. 
The other co-authors, who have expertise in XR, education, and human-AI interaction, did not participate in the CAE but joined the first and second authors to support group reflection and engage in the data analysis. 

\subsection{Reflective Journey}
Two researchers (T and X) independently used the Meta Ray-Ban AI Glasses over a one-month period (May 7–June 7, 2025). We selected this device for its integration of a state-of-the-art multimodal LLM (Llama 4) and its lightweight, socially acceptable form factor for everyday use. During the study period, Meta released an update enabling \textbf{Live AI\footnote{https://www.meta.com/help/ai-glasses/894093646030348/}}, allowing users to engage with the live-stream AI capabilities. As part of our CAE, we documented daily interactions through first-person diaries, capturing contextual details, motivations, emotional responses, and associated media \cite{lutz2019development}. Weekly meetings between the two researchers (one hour) and bi-weekly discussions with the broader team (one hour) supported ongoing reflection and consolidation of insights. All qualitative data, including diaries, meeting transcripts, and conversational logs, were analyzed using thematic analysis \cite{braun2012thematic}. We focus only on moments when the smart glasses leveraged their vision capabilities. The first author cleaned the transcripts, and both researchers independently conducted open-coding. Through joint discussions, we reconciled codes, identified axial categories, and organized them into higher-level themes that captured recurring successes and breakdowns in everyday interactions. Drawing from our distinct backgrounds (human–AI collaboration vs spatial computing) our analyses brought complementary lenses to interpreting the data. Broader team members also contributed alternative interpretations and counterexamples during bi-weekly meetings. We only report the patterns of conversational success and breakdowns in this poster as part of a larger project.

\section{Findings}

\subsection{Recurring Patterns of Conversational Success}
\begin{figure*}[t]
\centering
\includegraphics[width=\linewidth]{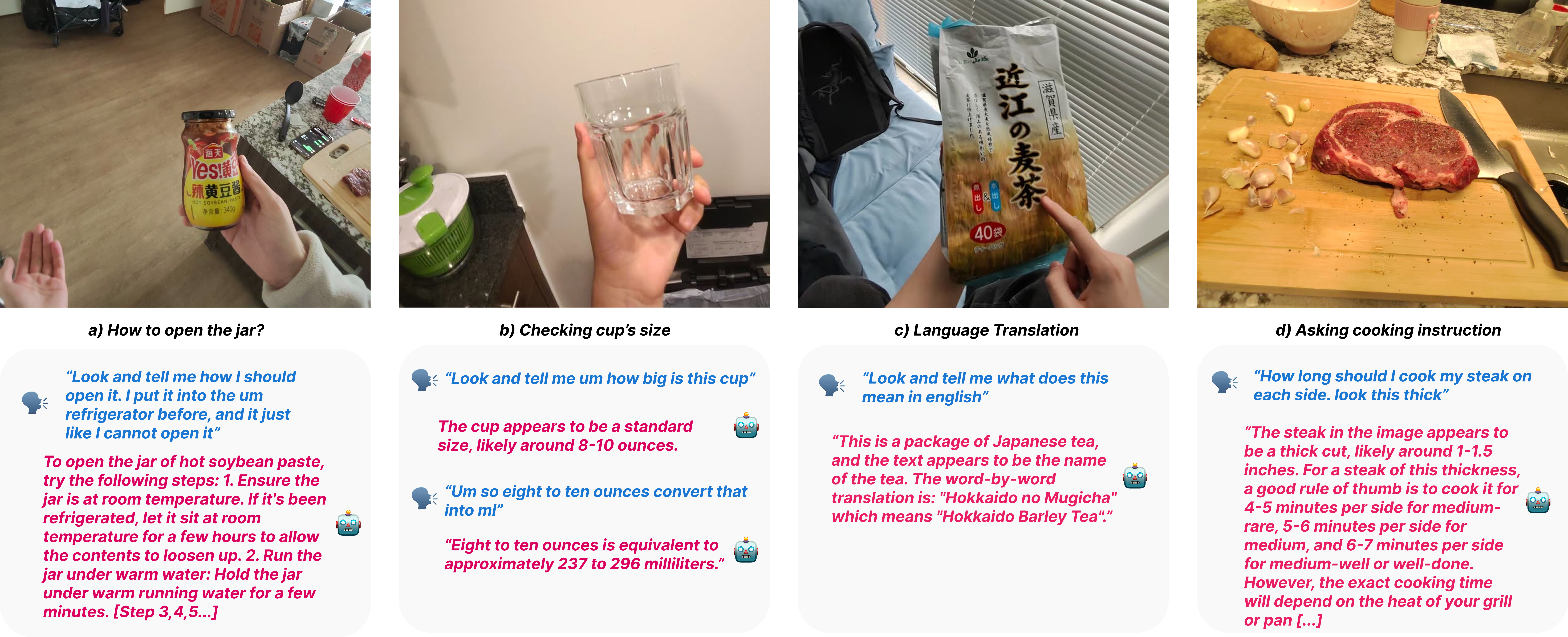}
\caption{This figure include four examples of successful conversational contexts regarding S1) Instant
Referential Problem-Solving and S2) Understanding Unfamiliar Knowledge.}
\Description{Four photographic examples of human–AI interaction scenarios paired with conversational transcripts. 
(a) A person holds a jar of hot soybean paste in a kitchen, with the AI providing step-by-step instructions to open it. 
(b) A person holds a clear glass cup, and the AI estimates its size in ounces and milliliters. 
(c) A person points to a package of Japanese barley tea with Japanese text, and the AI provides a translation into English. (d) A cutting board with a steak and garlic cloves, where the AI advises on cooking times per side based on steak thickness. Each example illustrates how smart glasses supported immediate task execution without requiring explicit references.}
\label{Successfully}
\end{figure*}

Through our collective reflections, we identified three recurring patterns of successful cases where smart glasses supported everyday uses. To begin with, we found that the pattern \textbf{S1) Instant Referential Problem-Solving} captures moments when participants used the smart glasses to quickly overcome practical obstacles in their routines using referential expressions. Rather than formulating detailed descriptions, we often used deictic terms such as “this” to point to physical objects in their environment. These interactions were typically brief and relied on the glasses’ visual capabilities to ground the referents and enable \textbf{immediate resolution without elaborate queries}. For instance, T recalled asking the AI for help with a frozen jar of paste (Fig\ref{Successfully}-a): \textit{“…I asked AI how to open the jar of paste [used ‘this’ in the actual query], because it was frozen. So I asked like, AI, how can I solve this question? And it just definitely worked. It said, Oh, you should […] put it into hot water, and just wait for like one hour…”} Similarly, X shared a related moment of relying on the glasses for real-time measurement (Fig~\ref{Successfully}-b): \textit{``…I just pointed at a glass cup and asked, How big is this cup? And it just said, standard 8 to 10 ounces… then converted it to milliliters. That’s what I needed.''}

Furthermore, we found that smart glasses could support users in \textbf{S2) Understanding Unfamiliar Knowledge} beyond straightforward problem-solving. In these moments, the AI was expected to provide interpretive or contextualized responses. Examples included translating labels while shopping (Fig.\ref{Successfully}-c) or clarifying cooking instructions (Fig.\ref{Successfully}-d). As S reflected, he often used the smart glasses for anything unfamiliar in his surroundings, noting that they felt \textit{``...just more intuitive and easier [than smartphones or voice assistants] when you’re walking—you don’t need to care about the audio, everything just works in place...''} Midway through our CAE, we found that the introduction of the Live AI feature expanded these possibilities by supporting multi-turn interactions for continuous knowledge acquisition. Although these interactions were not always dependable, we often experienced them as moments when the smart glasses acted as an adaptive tutor that could scaffold our understanding and reducing manuals sources.

Finally, we found that the smart glasses played a successful role in supporting our everyday \textbf{S3) Decision-Making}. This pattern typically arose when we faced multiple options and turned to the AI for guidance in choosing the most suitable path. One example emerged during T’s assembly of a utility cart. At a certain step, he was holding two nearly identical bases and asked the AI which one to use. As he recalled: \textit{``…here’s a really interesting case. There are two bases. I asked AI which I should use over here (the legs of the utility cart) … and AI told me you should use the one in your left hand … [And] the AI just taught me step by step, which I think is really cool, and I think they do remember …''} These successes highlighted how the smart glasses shaped our judgment in real time. Because the device was hands-free and always-on, these interactions felt lightweight and intuitive, enabling us to make choices without disrupting the flow of the original activity. 

When reflecting on our successful experiences, we also observed how the visual affordances of smart glasses shaped participants’ reflections throughout the CAE. Initially, we perceived this affordance as a convenient tool for verifying information. As T described, it was a “low-effort, quick solution” for him. Over time, however, we began to see the smart glasses as providing an “always-on spatial awareness.” For example, in the final meeting, T recounted his shopping experience when asking the AI where the garlic was: \textit{``…near the onions, other roots, vegetables. And it’s actually right. Because it [garlic] was in the corner, and the AI was just really good at telling me where it was…''}

\subsection{Recurring Patterns of Conversational Breakdowns}
\begin{figure*}[!t]
    \centering
\includegraphics[width=\linewidth]{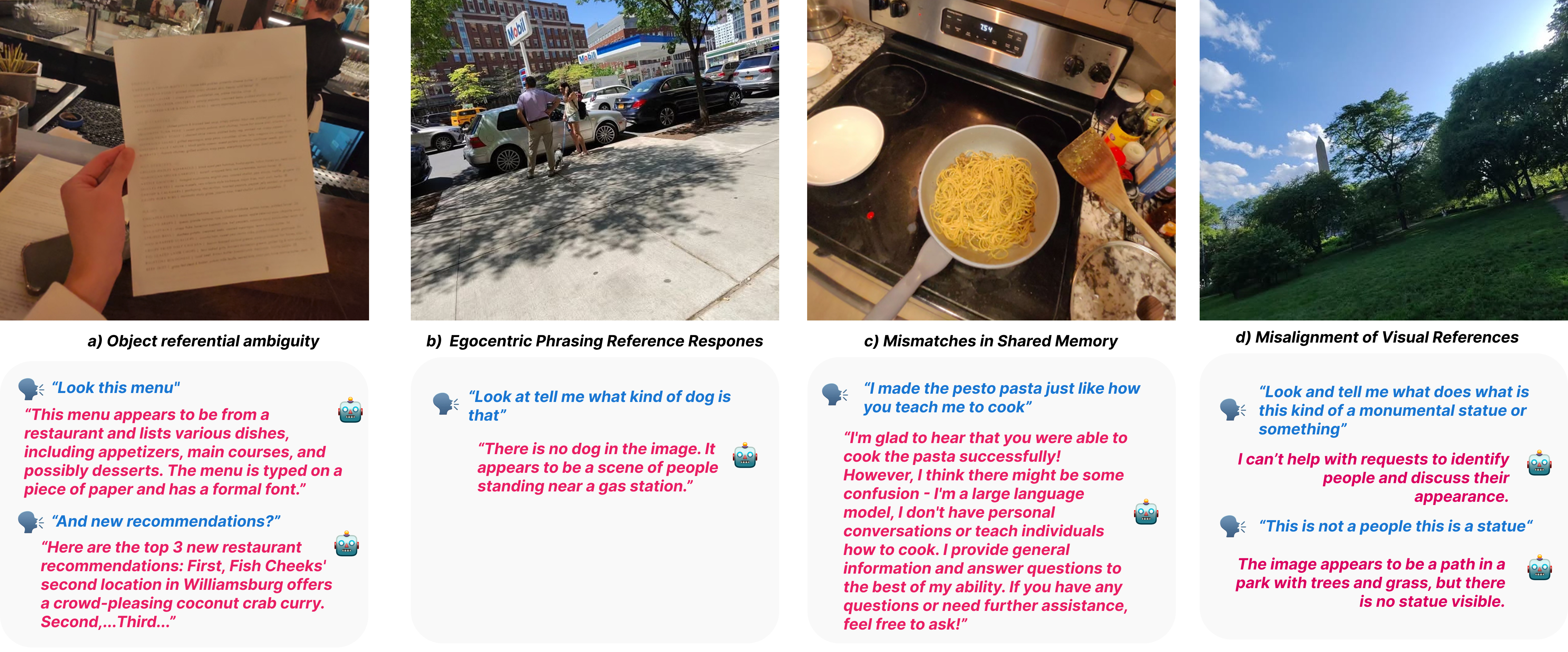}
    \caption{This figure includes four types of breakdown regarding response B1) Referential Rncoherence and B2) Conflict with Human Perception}
    \Description{Four photographic examples of human–AI interaction scenarios paired with conversational transcripts.
(a) A person holds a printed menu over a bar counter while the AI describes the menu and recommends restaurants.
(b) A city sidewalk scene near a Mobil gas station where a person asks about a dog, and the AI responds that no dog is present.
(c) A stovetop with cooked spaghetti in a pan, where the user mentions making pesto pasta and the AI clarifies it doesn’t provide personal cooking lessons.
(d) A wide-angle shot of a park with trees and a tall obelisk-like monument in the background; the user asks about a statue, but the AI states no statue is visible.}
    \label{referential}
\end{figure*}

By contrast, we identified three recurring types of conversational breakdowns when interacting with non-display smart glasses during everyday activities. One major issue is \textbf{B1) Referential Incoherence}, which includes object referential ambiguity (Fig.\ref{referential}-a) and mismatches in shared memory (Fig.\ref{referential}-c). Across our weekly reflections, we observed that many breakdowns occurred during task execution, often due to failures in maintaining referential coherence—in other words, such smart glasses were hard to follow users’ ongoing referential intent. These failures were then attributed to the system grounding responses solely on current visual input, difficulties handling intent transitions across tasks, and inconsistencies in the level of detail provided. For instance, X expressed his disappointment when locating a restaurant: \textit{``...But I wish it could be based on my current image, because at this moment the Chipotle is actually in front of me and could be shown in the image. So I think it could be answered just right in front of me...''} T also raised concerns about Live AI’s difficulty in supporting intent transitions across small, sequential tasks in a daily activity: \textit{``...After I left the potato section, I changed my mind and moved on to garlic. I wasn’t planning to use garlic with the potatoes, but the AI assumed the garlic was part of that same dish...''} 



\textbf{B2) Conflict with Human Perception} is another common breakdown. We find that such conflicts usually arise when system responses contradict users’ perceived visual recognition or their personal knowledge. Visual recognition failures often occur unexpectedly and frequently, caused by poor object quality, misalignment of visual references, or blurred recognition clarity (Fig. \ref{referential}-d). In such cases, responses often appear incomprehensible, leading users to feel confused and abandon further interaction out of nowhere. The underlying causes are difficult to diagnose until conversational logs are reviewed. These breakdowns may stem from unnoticed blur or objects being too small in the frame, yet users often do not realize this and instead terminate the interaction.

T and S both described how responses from devices sometimes directly conflicted with what they perceived (Fig.\ref{referential}-b). For example, S shared an experience when he wanted to identify a bird while touring in Central Park: \textit{S: ``I just asked, tell me what kind of bird it is. And it responded, there's no bird visible in the image.''} \textit{T: ``So how do you feel about that?''} \textit{S: ``I just feel, sort of, because I can see the bird. It responded very confidently, oh, there's no bird visible in the image... At least it could hedge, or it could suspect that this answer might be wrong, because I can see the bird...''} 
By contrast, when the responses of non-display smart glasses conflicted with users’ personal knowledge, the breakdown became even more complex, often undermining trust. For example, during the second week, T shared his navigation experience: \textit{``...I asked the AI if there is any coffee shop in [at this Airport], but the AI responded, `No,' in the search results. But I definitely know for sure the airport has a coffee shop..."} 

We also frequently felt \textbf{B3) Social Embarrassment}, which led to breakdowns in the presence of others. For instance, T recalled an incident while cooking: \textit{``...While I was talking with the AI, my girlfriend came out of her room and asked who I was talking to, since there are only two people in the apartment... I said, `my AI.' At that moment, she laughed, and I felt a bit awkward and embarrassed. I have no idea why, but I felt like a kid caught cheating by the teacher...''} Recalling this, X also shared his experience, leading to the following conversation: \textit{S: ``I was at Trader Joe's... I was holding a burrito and trying to ask if it could be heated in the microwave. But there were some people standing next to me. I decided not to do it, because I felt like, oh, there's already a label on it.''} \textit{T: ``I have to support you on that. I also feel that if there are people around me, I would definitely feel embarrassed to talk to the AI. It's just so weird, especially when others are talking with someone—it makes you feel awkward.''}

\textbf{B4) Voice-only Interaction Paradigm Limitations} of non-display smart glasses still lead to breakdowns similar to those of previous VAs, such as fixed interaction paradigms and difficulty understanding the system’s memory structure. S shared an example of how a conversation broke down during a query:\textit{ ``When I said ‘What, say it again,’ it weirdly lectured me about prefacing vision queries with ‘Hey Meta, look…’.”} T reflected on this when he shared another case about VA's limitation, saying: \textit{``...because the AI actually started four separate conversations... But I felt like I was just continuing the same dialogue history with the AI..."}. 

\section{What Makes Conversational Breakdowns and Successes Unique in Non-Display Smart Glasses?}

\begin{table*}[h]
\centering
\scriptsize
\caption{Comparison of Conversational Successes and Breakdown Patterns Between Prior VAs and Non-Display Smart Glasses}
\begin{tabularx}{\textwidth}{
    >{\raggedright\arraybackslash}p{2.5cm}
    >{\raggedright\arraybackslash}X
    >{\raggedright\arraybackslash}X
}
\toprule
& \textbf{Conversational Successes} & \textbf{Breakdown Patterns} \\
\midrule
\textbf{Prior VAs}
& Quick responses to simple queries; routine commands perform predictably \cite{cowan_what_2017,ammari_music_2019,bentley_understanding_2018}.
& \textbf{Voice-only limitations:} Misrecognized intent, ambiguous phrasing, lack of context. Errors in low-stakes settings \cite{mahmood2025user,porcheron_voice_2018}. \\
\midrule
\textbf{Non-Display Smart Glasses (Ours)}
& \textbf{S1:} Instant referential problem-solving with real-world scenes. \textbf{S2:} Understanding unfamiliar knowledge and \textbf{S3:} supporting decision-making during tasks
& \textbf{B1:} Referential incoherence and \textbf{B2:} Conflict with human visual perception during the task. \textbf{B3:}  Public social embarrassment in everyday context. \textbf{B4:} Voice-only limitation still exists. \\
\bottomrule
\end{tabularx}
\label{tab:comparison_horizontal}
\end{table*}

Our findings reveal the patterns of conversational success and breakdown in everyday use of non-display smart glasses. Compared to patterns observed in traditional and LLM-based voice assistants, successes were indeed centered around simple query interactions. However, smart glasses stood out in their ability to provide situated and real-time guidance with blurred referents in ways that static voice assistants rarely achieve (which typically offer quick answers to static queries) \cite{cowan_what_2017, ammari_music_2019, bentley_understanding_2018}. This capability suggests that smart glasses can support everyday tasks as an `always-on tutor,’ aligning with prior work on context-sensitive wearable AI \cite{arakawa_prism-q_2024, cai_aiget_2025, ning_aroma_2025}, while extending it by demonstrating how voice-only interaction (as opposed to visual overlays) can scaffold decision-making in everyday life.

Conversational breakdowns were uniquely amplified by the embodied affordances of non-display smart glasses. Referential problems (e.g., this one,' the item in my left hand’) were not merely linguistic challenges of visual grounding as observed in prior voice assistant studies \cite{mahmood_voice_2025, chan2025mango}. Instead, it was the sustained failure in maintaining referential coherence that proved critical. In these moments, understanding how such smart glasses continue to track and align with user intent emerged as more important than resolving the initial referential failure. Moreover, the visual capability of smart glasses introduced new challenges. Users expected the device to `see what they see,’ but discrepancies in clarity, distance, and perspective often meant that when smart glasses confidently responded in ways that contradicted visually obvious realities (e.g., denying the presence of a visible bird), users experienced strong emotional reactions and a loss of trust. Unlike smart speakers, where vague or incomplete answers can be overlooked, breakdowns in smart glasses directly undermine the premise of shared perception. A further distinctive dimension was the social and cultural context of use. In contrast to smart speakers used in private homes \cite{ammari_music_2019, chan2025mango, yang_talk2care_2024, porcheron_voice_2018}, smart glasses are worn in public, rendering breakdowns socially consequential. Users often abandoned queries in the presence of others to avoid embarrassment, echoing findings from recent studies on wearables \cite{zhang_through_2025, lee_sensible_2025}. While referential incoherence is a common problem in VAs, the consequences of and solutions to conversational breakdowns change in voice-only smart glasses are more likely to influence users’ social behavior and pace of action in public context. We elaborate on these comparisons between VAs and non-display smart glasses in Table~\ref{tab:comparison_horizontal}.

\section{Conclusion and Limitation}

Our poster contributes to the HCI community by examining current conversational breakdowns and successes with non-display smart glasses. 
As an exploratory study, our findings are not intended to be broadly generalizable; instead, we surface initial themes that merit further investigation in larger-scale, more diverse, and methodologically rigorous research. We acknowledge that our positionalities, as both users and investigators, have shaped the experiences we analyzed and the interpretations we drew. As both participants possess technical familiarity with AI systems, their awareness of breakdowns and recovery strategies may differ from novice users. Future work should expand to include a broader and more diverse set of users to capture variations in everyday practices and cultural norms.



\bibliographystyle{ACM-Reference-Format}
\bibliography{main}

\end{document}